\documentclass[11pt]{article}

\usepackage{amsmath,amsfonts,amsbsy,amssymb,array,accents,dsfont}
\usepackage{enumerate,array,latexsym,graphicx,mathrsfs,verbatim,psfrag}
\usepackage{bm} 
\usepackage[normalem]{ulem}
\usepackage{booktabs} 
\usepackage[usenames]{color}
\usepackage[utf8]{inputenc}
\usepackage[english]{babel}
\usepackage{fancybox}

\usepackage{datetime}

\usepackage[nosort]{cite}
\usepackage{chngpage} 

\usepackage{enumitem}
\setlist{itemsep=0pt}

\usepackage[colorlinks=true,      linkcolor=darkblue,      urlcolor=darkblue,      
            filecolor=darkblue,      citecolor=darkblue,       pdfstartview=FitH,     
						pdfpagemode=UseNone,      bookmarksopen=true]{hyperref}  
\usepackage[all]{hypcap}     

\newcommand{\captionfonts}{\small}
\makeatletter  
\long\def\@makecaption#1#2{%
  \vskip\abovecaptionskip
  \sbox\@tempboxa{{\captionfonts #1: #2}}%
 \ifdim \wd\@tempboxa >\hsize
    {\captionfonts #1: #2\par}
  \else
    \hbox to\hsize{\hfil\box\@tempboxa\hfil}%
  \fi
  \vskip\belowcaptionskip}
\makeatother   

\DeclareMathSymbol{\medhatsym}{\mathord}{largesymbols}{"62} 

\DeclareMathSymbol{\medtildesym}{\mathord}{largesymbols}{"65}

\newcommand{\comm}[1]{} 

\def\IS{\mathbb{S}}
\def\IT{\mathbb{T}}

\def\({\left(}
\def\){\right)}
\def\[{\left[}
\def\]{\right]}

\def\sst{\scriptscriptstyle}

\def\coeff#1#2{{\textstyle \frac{#1}{#2}}}

\def\One{{\hbox{ 1\kern-.8mm l}}}

\def\barray{\begin{array}}
\def\earray{\end{array}}
\def\be{\begin{equation}}
\def\ee{\end{equation}}
\def\bea{\begin{eqnarray}}
\def\eea{\end{eqnarray}}
\def\bal{\begin{align}}
\def\eal{\end{align}}

\def\mR{\mathbb{R}}

\numberwithin{equation}{section} 

\makeatletter
\g@addto@macro\bfseries{\boldmath}
\makeatother

\definecolor{cardinal}{rgb}{0.6,0,0}
\definecolor{darkgreen}{rgb}{0,0.4,0}
\definecolor{purple}{rgb}{0.5, 0, 0.5}
\definecolor{golden}{rgb}{0.92, 0.7, 0}
\definecolor{midnight}{rgb}{0, 0, 0.5}
\definecolor{darkblue}{rgb}{0, 0, 0.8}

\def\Neql#1{{\cal N}\!=\!{#1}}
\def\coeff#1#2{\relax{\textstyle {#1 \over #2}}\displaystyle}

\def\ZZ{\mathds{Z}}

\def\cB{{\cal B}}

\def\cF{{\cal F}}

\def\cK{{\cal K}}

\def\cM{{\cal M}}

\def\cP{{\cal P}}

\def\cS{{\cal S}}

\def\nBPS#1{$\frac{1}{#1}$-BPS}

\def\bbS{\mathbb{S}}

\def\sst#1{\scriptscriptstyle{#1}}

\newcommand{\bn}{\ensuremath{j}}

\newcommand{\np}{\ensuremath{n_{\mathrm{\sst{P}}}}}

\newcommand{\subsubsectionmod}[1]{
\refstepcounter{subsubsection}
\subsubsection*{\thesubsubsection ~~  #1}
}

\begin{document}


\begin{flushright}
IPHT-T17/134\\
\end{flushright}

\vspace{14mm}

\begin{center}

{\huge \bf{Integrability and Black-Hole}} \medskip

{\huge \bf{Microstate Geometries}}

\vspace{13mm}

\centerline{{\bf Iosif Bena$^1$,  David Turton$^1$, Robert Walker$^{2}$ and Nicholas P. Warner$^{2,3}$}}
\bigskip
\bigskip
\vspace{1mm}

\centerline{$^1$\,Institut de Physique Th\'eorique,}
\centerline{Universit\'e Paris Saclay, CEA, CNRS, }
\centerline{Orme des Merisiers,  F-91191 Gif sur Yvette, France}
\bigskip
\centerline{$^2$\,Department of Physics and Astronomy,}
\centerline{University of Southern California,} \centerline{Los
Angeles, CA 90089-0484, USA}
\bigskip
\centerline{$^3$\,Department of Mathematics,}
\centerline{University of Southern California,} \centerline{Los
Angeles, CA 90089, USA}

\vspace{4mm}

{\small\upshape\ttfamily iosif.bena @ ipht.fr, ~david.turton @ ipht.fr,} \\ 
 {\small\upshape\ttfamily  walkerra @ usc.edu, ~warner @ usc.edu} \\

\vspace{10mm}

\textsc{Abstract}

\begin{adjustwidth}{17mm}{17mm} 
 %
\vspace{3mm}
\noindent
We examine some recently-constructed families of asymptotically-AdS$_3 \times \bbS^3$ supergravity solutions that have the same charges and mass as supersymmetric D1-D5-P black holes, but that cap off smoothly with no horizon. These solutions, known as superstrata, are quite complicated, however we show that, for an infinite family of solutions, the null geodesic problem is completely integrable, due to the existence of a non-trivial conformal Killing tensor that provides a quadratic conservation law for null geodesics. This implies that the massless scalar wave equation is separable. For another infinite family of solutions, we find that there is a non-trivial conformal Killing tensor only when the left-moving angular momentum of the massless scalar is zero. We also show that, for both these families, the metric degrees of freedom have the form they would take if they arose from a consistent truncation on $\bbS^3$ down to a $(2+1)$-dimensional space-time. We discuss some of the broader consequences of these special properties for the physics of these black-hole microstate geometries. 
%
\end{adjustwidth}

\end{center}


\thispagestyle{empty}

\newpage


\baselineskip=14pt
\parskip=2pt

\tableofcontents


\baselineskip=15pt
\parskip=3pt

\section{Introduction}
\label{Sect:introduction}
 
The last year has seen a significant breakthrough in the construction of microstate geometries \cite{Bena:2016ypk,Bena:2017geu}.  In particular, microstate geometries corresponding to  five-dimensional, three-charge, supersymmetric black holes with arbitrarily small angular momenta have been constructed. These solutions are horizonless and smooth and have an arbitrarily-long BTZ-like throat that interpolates between an AdS$_3 \times \bbS^3$ asymptotic region and a long very-near-horizon AdS$_2$ region. Deep inside the AdS$_2$ region, the throat caps off smoothly just above where the black-hole horizon would be~\cite{Bena:2016ypk}.

In the  D1-D5-P frame these microstate geometries, known as ``superstrata'' \cite{Bena:2011uw,Bena:2015bea}, have been proposed as holographic duals of specific families of pure states of the D1-D5 CFT, involving particular left-moving momentum-carrying excitations~\cite{Bena:2016ypk}\footnote{A subset of these microstate geometries can be mapped to excitations of the MSW string that carry momentum and angular momentum, via a sequence of solution-generating transformations and string dualities~\cite{Bena:2017geu}.}, with charges in the regime of parameters in which a large BPS black hole exists. Hence these solutions correspond to microstates of a black hole with a macroscopic horizon. 
The momentum excitations may be thought of as creating the long AdS$_2$ black-hole-like throat; in the full solution these momentum excitations are located deep inside that throat, and support its macroscopic size. As one descends the AdS$_2$ throat, to an excellent approximation it is almost identical to a black hole throat until near the bottom, where one encounters the momentum excitations, before the geometry caps off smoothly. 

Given that we have a holographic understanding of these solutions, and that the proposed dual CFT states live in the same ensemble as the states that give rise to the black hole entropy, it is important to identify the physical consequences of the fact that these solutions lack horizons. One expects the classical black hole solution to give a thermodynamic coarse-grained description of the physics, and for the bulk description of typical black hole microstates to give the correct fine-grained description of the physics. For simple physical processes, typical states should reproduce the thermodynamic coarse-grained physics (see e.g.~\cite{Mathur:2012jk,Mathur:2013gua}), and should also give rise to novel physics where the thermodynamic description breaks down. Hence, we would like to understand how, for example, horizonless geometries scatter and absorb incoming particles, and how this differs from the classical black hole result. The states we will study are still somewhat atypical, and will have interesting differences from the corresponding classical black hole; we hope the present study will inform future studies of progressively more typical microstates.

Superstratum solutions are parameterized by arbitrary functions of (at least) two variables~\cite{Bena:2015bea}. Generic superstratum solutions depend on all but one of the coordinates in six dimensions, and hence may appear complicated and somewhat intimidating to the uninitiated.  However, in this paper we will show that two particular asymptotically-AdS$_3 \times \bbS^3$  families, each parameterized by one positive integer, have much simpler physics than may have been expected. One of these families, which we will call the  $(1,0,n)$ family, has a separable wave equation and  a conformal Killing tensor\footnote{This provides a quadratic conserved quantity for null geodesics.}. This implies that the equations for null geodesics in these geometries are completely integrable: there is a complete set of conserved quantities, that are linear or quadratic in velocities.  
Related work on geodesic integrability in (two-charge) black hole microstate geometries and D-brane metrics can be found in~\cite{Chervonyi:2013eja,Chervonyi:2015ima}.

 We will also show that the metrics of the $(1,0,n)$ family, and of another family that we will call the  $(2,1,n)$ family, can be re-written in a form that they would take if they arose from consistent truncations on $\IS^3$ to $(2+1)$-dimensions.  By this, we mean the following: we can write each metric in terms of a (deformed) $\IS^3$ fibration over a three-dimensional base, $\cK$, where the metric on $\cK$ depends only on the coordinates on $\cK$.  Moreover, the fibration and warp factors conform to the standard KK Ansatz for vector fields and Einstein gravity on $\cK$.  

In this paper we will restrict our attention to the metric degrees of freedom, and we postpone a full analysis of the existence, or otherwise, of a complete consistent truncation to future work.  Such an investigation would require the inclusion of the six-dimensional tensor gauge fields in the consistent truncation Ansatz, building upon the results of \cite{Cvetic:2000dm, Deger:2014ofa} to include more six-dimensional tensor multiplets. The purpose of the present work is to elucidate the metric  structure and how it takes the form it would take if it came from a consistent truncation, since this is a remarkably strong constraint upon its structure. The metrics of the  $(1,0,n)$ and $(2,1,n)$ families also have several isometries, such that the reduced metric on $\cK$, and the KK fields of the fibration, {\it depend only on one coordinate} (which asymptotically becomes the radial coordinate of the AdS$_3$). Thus, the analysis of these metrics can, in principle, be carried out entirely using three-dimensional gravity coupled to vector fields.

Unlike in the $(1,0,n)$ family, the wave equation is not generically separable in the $(2,1,n)$ family, and the general geodesic problem does not appear to be completely integrable. However, these microstate geometries come very close to having these properties:  For waves or geodesics that have zero SU$(2)_L$ angular momentum on the $\bbS^3$, one {\it does} have separability of the wave equation and complete integrability of the null geodesic equations.  

As we will discuss, complete integrability of geodesics may be both a blessing and a curse.  In particular, completely integrable systems have highly restricted spectra and limited scattering behavior, and thus may not reveal some of the interesting physics of generic microstate geometries. The complete set of conservation laws lead us to suspect that the $(1,0,n)$ geometries will quickly eject infalling particles, and hence will not reproduce the expected black-hole thermodynamic behavior. On the other hand, the absence of integrability in the $(2,1,n)$ family will allow for more complicated dynamics, and possibly the trapping of incoming particles, which is a step closer to the behavior one expects from typical microstates. In a sense, the $(2,1,n)$ family provides an ideal setting for investigating more complicated dynamics: one can probe the behavior of the non-integrable geodesics by doing perturbation theory for waves and geodesics about the integrable ($j_L=0$) ones.  Thus one can probe quite non-trivial scattering properties in a controlled expansion.

There are three recent, seemingly unrelated, lines of investigation to which our results should be relevant. The first is an argument that supersymmetric microstate geometries should exhibit a non-linear instability~\cite{Eperon:2016cdd} (see also~\cite{Keir:2016azt,Marolf:2016nwu,Eperon:2017bwq}). This proposed non-linear instability is related to the existence of stably trapped null geodesics deep inside the core of microstate geometries, however so far this has only been explicitly analyzed for very symmetric geometries, none of which have charges and angular momenta corresponding to a black hole with a large horizon area. Our results should help elucidate whether or not this proposed non-linear instability is an artifact of very symmetric microstate geometries.

The second line of investigation is the late-time behavior of correlation functions in the D1-D5 CFT, and its connection to quantum chaos \cite{Maldacena:2015waa}. One expects chaotic systems to exhibit late-time fluctuations that come from their underlying microscopic description, and that are not visible in the thermodynamic approximation. Hence, one expects typical bulk microstates to give rise to late-time fluctuations that are not visible in the classical black-hole solution. For a set of two-charge black hole microstates these fluctuations have been computed in the dual CFT \cite{Balasubramanian:2005qu,Balasubramanian:2016ids}, but the corresponding computations in the bulk are beyond the capability of present technology. We hope that our results will open the way for a new testing ground for these questions.

The third line of investigation is the computation of four-point functions of two heavy and two light operators; for a few examples, see~\cite{Fitzpatrick:2015zha, Hijano:2015qja, Anous:2016kss, daCunha:2016crm, Galliani:2016cai,Galliani:2017jlg, Chen:2017yze}.  Such calculations can be done in the CFT and can be matched to the light-light two-point function computed holographically in the microstate geometry dual to the heavy state \cite{Galliani:2016cai,Galliani:2017jlg}. So far, the bulk calculation has only been done in microstate geometries that are dual to very special heavy states (a particular set of R-R ground states and spectral flows thereof~\cite{Giusto:2012yz}), essentially because of the technical difficulty of solving the wave equation. The integrability of geodesics and the separability of wave operators in the infinite families of microstate geometries we consider should simplify the calculation of the two-point functions  and vastly enhance the number of two-heavy-two-light four-point functions that can be computed in the bulk. Moreover, since there is an explicit proposal for the CFT states dual to the microstate geometries we study~\cite{Bena:2016ypk}, these four-point functions could be compared to those computed in the CFT. Furthermore, since our geometries can have long black-hole-like AdS$_2$ throats and closely resemble black holes far from the cap region, these four-point functions should shed light on how unitarity is restored when replacing the black hole horizon with a fuzzball. 

This paper is organized as follows.
In Section \ref{sec:DimRedGen} we describe in more detail the special properties of the metrics in the  $(1,0,n)$ and  $(2,1,n)$ families of microstate geometries.      In Section \ref{sec:SingleMode} we give a brief description of these families of BPS solutions.  
In Section \ref{sec:Special-properties} we give the details of the metrics of the $(1,0,n)$ and  $(2,1,n)$ families, separate variables in the wave equation, and describe the conformal Killing tensor.  Finally, in Section~\ref{sec:Disc} we discuss the important features of the geometries described in this paper, and further discuss the implications of our results.

\section{Dimensional reduction, separability and  Killing tensors} 
\label{sec:DimRedGen}

We work in Type IIB string theory on $\cM^{4,1}\times \bbS^1\times \cM$, where $\cM$ is either $\IT^4$ or K3. The circle, $\bbS^1$, is taken to be macroscopic and is parameterized by the coordinate $y$, with radius $R_y$:
\begin{equation}
y ~\sim~ y \,+\, 2 \pi R_y  \,. 
\label{yperiod}
\end{equation}
We consider a bound state of D1-branes wrapped on $\bbS^1$, D5-branes wrapped on $\bbS^1\times \cM$, and momentum P along $\bbS^1$.
The internal manifold, $\cM$, is taken to be microscopic, and we assume that all fields are independent of  $\cM$.  
Upon dimensional reduction on $\cM$, one obtains a theory whose low-energy limit is six-dimensional $\Neql1$ supergravity coupled to two (anti-self-dual) tensor multiplets. This theory contains all fields expected from the study of D1-D5-P string world-sheet amplitudes \cite{Giusto:2011fy}. The system of equations describing all \nBPS{8} solutions of this theory was found in  \cite{Giusto:2013rxa}; it is a generalization of the system discussed in \cite{Gutowski:2003rg,Cariglia:2004kk}.  Most importantly, this BPS system can be greatly simplified, and largely linearized \cite{Bena:2011dd}.
For supersymmetric solutions the six-dimensional metric is well known to take the form \cite{Gutowski:2003rg}:
\begin{equation}
ds_6^2 ~=~    -\frac{2}{\sqrt{\cP}} \, (dv+\beta) \big(du +  \omega + \tfrac{1}{2}\, \mathcal{F} \, (dv+\beta)\big) 
~+~  \sqrt{\cP} \, ds_4^2(\cB) ~\equiv~ g_{MN} dz^M d z^N\,, \label{sixmet}
\end{equation}
where we take
\begin{equation}
  u ~=~  \coeff{1}{\sqrt{2}} (t-y)\,, \qquad v ~=~  \coeff{1}{\sqrt{2}}(t+y) \,. \label{tyuv-1}
\end{equation}
Supersymmetry requires that all fields be independent of $u$, but generic supersymmetric solutions can depend upon all the other coordinates.

Upon taking the AdS/CFT decoupling limit~\cite{Maldacena:1997re}, one obtains asymptotically AdS$_3 \times \bbS^3$ solutions. We will work exclusively in the decoupling limit throughout this paper.
We shall study solutions whose tensor fields have explicit dependence on $v$, as well as on the $\bbS^3$.
These solutions are known as ``superstrata''~\cite{Bena:2011uw,Bena:2015bea,Bena:2016agb,Bena:2016ypk,Bena:2017geu}.
In the solutions we study, the metric, $ds_4^2$, on the four-dimensional base, $\cB$, is flat and we write it in the standard bipolar form:
\begin{equation}
ds_4^2 ~\equiv~ \tilde g_{ab} d z^a d z^b  ~=~ \Sigma\, \bigg( \frac{dr^2}{(r^2 + a^2)} ~+~ d \theta^2 \bigg)  ~+~ (r^2 + a^2) \sin^2 \theta \, d\varphi_1^2 ~+~ r^2  \cos^2 \theta \, d\varphi_2^2   \,,
\label{basemet}
\end{equation} 
where 
\begin{equation}
\Sigma ~\equiv~ (r^2 + a^2 \cos^2 \theta) \,.  \label{Sigdefn}
\end{equation}
At infinity, the sets of coordinates $(u,v,r)$ and $(\theta, \varphi_1, \varphi_2)$ parametrize AdS$_3$ and $\bbS^3$ respectively.    The superstratum solutions that we consider were constructed in \cite{Bena:2016ypk}, and they have the property that the tensor fields depend explicitly on a single linear combination of $v$, $\varphi_1$ and $\varphi_2$. We thus refer to them as {\it single-mode superstrata}.  In these asymptotically AdS$_3 \times \bbS^3$ solutions, this phase-dependence cancels in the energy-momentum tensor, and hence in the metric\footnote{Upon completing these solutions to asymptotically $\mR^{1,4}\times$S$^1$ solutions, the metric depends explicitly on the linear combination of $v$, $\varphi_1$ and $\varphi_2$~\cite{BGMRSTW-II}.}.  Thus the metric has isometries not just along  along $u$ (as required by supersymmetry) but also along $v$, $\varphi_1$ and $\varphi_2$.  In this paper we shall exploit these enhanced symmetries and examine the remaining, highly non-trivial dependence  on $(r,\theta)$.

Our first goal is to re-write the general six-dimensional metric as a fibration of the compact three-manifold, $\cS$, described by $(y^1,y^2,y^3) \equiv (\theta, \varphi_1, \varphi_2)$, over a base, $\cK$, parametrized by $(x^1,x^2,x^3)  \equiv (u,v,r)$.  Specifically, we re-cast (\ref{sixmet}) in the following form:
\begin{equation}
ds_6^2  ~\equiv~ g_{MN} dz^M d z^N~=~     \Omega^{-2} \,  \hat g_{\mu \nu} dx^\mu dx^\nu  ~+~  h_{i j} (dy^i  + {B^i}_\mu dx^\mu) (dy^j  + {B^j}_\nu dx^\nu)   \,, \label{fiberform}
\end{equation}
where $\hat g_{\mu \nu}$ and $h_{i j}$ are viewed as metrics on $\cK$ and $\cS$ respectively, and where $\Omega$ is defined to be the volume of $\cS$ divided by  the volume of the $\bbS^3$ to which $\cS$ limits at infinity: 
\begin{equation}
\Omega ~\equiv~     \frac{\sqrt{det(h_{ij})}}{\sqrt{det(h_{ij})}|_{r \to \infty}} \,. \label{Omdefn}
\end{equation}
In a general BPS solution, $\hat g_{\mu \nu}$, $h_{i j}$, ${B^i}_\mu$ and $\Omega$ can depend on all the coordinates, except $u$. 

It is  convenient to define the metrics:
\begin{equation}
ds_{1,2}^2 ~\equiv~     \hat g_{\mu \nu} dx^\mu dx^\nu  \,, \qquad ds_3^2 ~\equiv~      h_{i j} dy^i dy^j \,. \label{basefibermet}
\end{equation}
At infinity, $ds_{1,2}^2$ is asymptotic to the metric on AdS$_3$, and $ds_3^2$ is asymptotic to the metric on $\bbS^3$.  It is also useful to observe that one can invert  the form of  $g_{MN}$ in (\ref{fiberform}) explicitly:
\begin{equation}
g_{MN} =  \begin{pmatrix}
\Omega^{-2} \,\hat g_{\mu \nu} + h_{k m} \, {B^k}_\mu {B^m}_\nu&  {B^k}_\mu \,h_{kj} \\
h_{ik} \,{B^k}_\nu  &  h_{ij}  
\end{pmatrix}  \,,   \quad
g^{MN} =\Omega^{2} \, \begin{pmatrix}
\hat g^{\mu \nu}  & -\hat g^{\mu \rho} \, {B^j}_\rho  \\
- {B^i}_\rho \hat g^{\rho \nu} \,   &  { \Omega^{-2} } h^{ij}  + \hat g^{\rho \sigma} {B^i}_\rho {B^j}_\sigma
\end{pmatrix}   . \label{matrixforms}
\end{equation}
In particular, we note that the inverse metric on the internal space, $\cS$, is a non-trivial combination of the vector fields, ${B^i}_\mu$, and the metrics $\hat g_{\mu \nu}$ and $h_{ij}$.

The warp factor, $\Omega^{-2}$,  in front of $ds_{1,2}^2$ in (\ref{fiberform}) is precisely the factor needed for the dimensional reduction from six dimensions down to the three-dimensional space time, $\cK$.  To be more specific, this is the warp factor needed to reduce the six-dimensional Einstein action down to the three-dimensional Einstein action for $ \hat g_{\mu\nu}$ on $\cK$.   In general, attempting to perform such a dimensional reduction is of course not very useful, because $\hat g_{\mu \nu}$ in (\ref{fiberform}) will typically depend upon the $y^i$.

However, the $(1,0,n)$ and $(2,1,n)$ single-mode families of BPS geometries both have the property that:
\begin{itemize}
\item[(i)]  The metric $\hat g_{\mu \nu}$ is {\it only} a function of $r$.
\end{itemize}
Moreover, the $(1,0,n)$ family also has the following remarkable feature:
\begin{itemize}
\item[(ii)]  On  $ds_6^2$,  the massless wave equation and the  Hamilton-Jacobi equation for null geodesics are separable.  The ``massive''  wave equation on $ds_6^2$ is also separable if the mass term is induced from a mass as seen by the the (2+1)-dimensional metric $ds_{1,2}^2$.
\end{itemize}
Finally, it is elementary to verify that the following is true for all metrics of the form  (\ref{sixmet}):
\begin{itemize}
\item[(iii)]  If one computes  $\sqrt{-g} \, g^{MN}$ for the six-dimensional metric, then the components of this along the base defined by $(r,\theta,\varphi_1, \varphi_2 )$ are identical to the components of   $\sqrt{-\tilde g} \, \tilde g^{ab}$, where $\tilde g^{ab}$ is the metric defined in (\ref{basemet}).
\end{itemize}
%

Property (i), combined with equations (\ref{fiberform}) and (\ref{Omdefn}),  means that the complete six-dimensional metric has the form it would take if it arose from a consistent truncation to three-dimensional physics on $\cK$. This three-dimensional geometry is furthermore {\it determined entirely by functions of $r$ alone.}

Property (ii), the separability of the massless wave equation and of the massless Hamilton-Jacobi equation  for null geodesics, implies the existence of a ``hidden symmetry'': there is another quadratic conserved quantity for the null geodesic equation~\cite{Carter:1968rr,Carter:1968ks}\footnote{See also the recent review on geodesic integrability in black-hole backgrounds~\cite{Frolov:2017kze}.}. 
That is, in addition to the usual conserved quantities along geodesics, there is a conformal Killing tensor, $\xi_{MN}$, for which one has:
\begin{equation}
\frac{D}{D \lambda} \bigg(\xi_{MN} \, \frac{dz^M}{d \lambda} \, \frac{dz^N}{d \lambda} \bigg)  ~=~    F(\lambda)\,  \bigg(g_{MN} \, \frac{dz^M}{d \lambda} \, \frac{dz^N}{d \lambda} \bigg) \,, \label{confKT}
\end{equation}
for some function $F(\lambda)$ along each geodesic.  In particular, the right-hand side vanishes for null geodesics.  Since the metrics we are considering have four Killing vectors, a conformal Killing tensor and the usual conserved quadratic form from the metric, the null geodesic problem is completely integrable:  It has four conserved momenta that are linear in velocities and two quadratic ``energies,'' one involving $v_r \equiv  \frac{dr}{d \lambda}$ alone and the other involving $v_\theta \equiv  \frac{d\theta}{d \lambda}$ alone.

Properties (ii) and (iii) together mean that not only is the massless wave equation separable, but its separability properties are precisely those of the flat-space base metric written in spherical bipolars (\ref{basemet}).  In particular, the angular modes on $\cS$ are elementary:  they are simply the standard spherical harmonics on a round $\bbS^3$!  Therefore the solutions of the massless wave equation have an expansion in terms of functions of $r$ alone, multiplied by Jacobi polynomials in  $\cos^2 \theta $.  Thus, most of the interesting physics is encoded in the radial equation and in the functions of $r$ alone that define $\hat g_{\mu \nu}$.
 
Finally, property (iii) suggests a rather interesting conjecture arising from  the general lore of consistent truncation  of supergravity theories compactified on spheres.  Typically, purely internal, higher-dimensional excitations reduce to scalar fields in the lower-dimensional theory.  Conversely, one of the most complicated aspects of obtaining  ``uplift'' formulae for consistent truncations is the way in which lower-dimensional scalars encode the details of the higher-dimensional fields.  For maximally supersymmetric theories on spheres, there is now a large literature on this, but one of the earliest breakthroughs were the metric uplift formulae \cite{deWit:1984nz,Cvetic:2000dm,Pilch:2000ue,Deger:2014ofa}.  These formulae gave complete and explicit expressions, in terms of the lower-dimensional scalar fields, for the inverse metric projected onto the internal manifold.  One of the simple consequences of this formula is that if the inverse metric retains its original round form, then it means that  the lower-dimensional scalars are essentially trivial.

This piece of lore suggests that in the solutions we are considering, there are no fundamental, lower-dimensional scalar excitations arising from the six-dimensional metric:  All the internal physics is encoded in the vector multiplets that descend from the ${B^i}_\mu$.

There are several caveats that come with this comment. First, we have not analyzed the tensor gauge fields to determine whether or not a complete, consistent truncation containing the above solutions exists.  Since these tensor fields have non-trivial dependence on the $\bbS^3$ directions, they will descend to massive fields in three dimensions. 

Moreover, while important results have been obtained in \cite{Deger:2014ofa}, the  general consistent truncation formulae have not been  established for the $\bbS^3$ compactifications considered here. In addition, the formulae that have been established in other reductions are based on massless vector fields that descend through Killing vectors on the sphere.  For the tensor fields and for some of the metric components in the $(2,1,n)$ family, we will need to allow more general classes of fields, ${B^i}_\mu$, in which the internal components involve higher harmonics, yielding massive vector fields on $\cK$. As we shall comment on below, it is not clear how restricted an ansatz might be necessary in order to obtain a consistent truncation.

 It is also important to note that, from the three-dimensional perspective, Abelian vector fields can trivially be re-written as scalars.  On the other hand, there are subtleties in doing this for non-Abelian fields and with off-shell supermultiplet structure  (see, for example, \cite{Aharony:1997bx}) and so the idea that there are only excitations of three-dimensional vector fields, descending from metric modes on $\bbS^3$, may be given some more precise formulation.  The bottom line is that the supergravity lore on consistent sphere truncations suggests that property (iii) might imply that the only  degrees  of freedom that are being activated in our solutions are the vector fields encoded in  ${B^i}_\mu$ and that there are no other independent shape modes or lower-dimensional scalars coming from the six-dimensional metric.

\section{Single-mode superstrata} 
\label{sec:SingleMode}

In this section we review the construction of superstrata, before focusing on the set of such solutions that involve a single-mode excitation. This will provide some background and allow us to set up notation to be used when we present our main results in the next section. 

\subsection{D1-D5-P Superstrata}
\label{ss:SSgeom}

The superstrata constructed to date have been obtained~\cite{Bena:2015bea,Bena:2016agb,Bena:2016ypk} by adding momentum waves to the background of the circular supertube~\cite{Balasubramanian:2000rt,Maldacena:2000dr,Mateos:2001qs, Lunin:2001fv,Emparan:2001ux}.   The starting point is therefore to take the vector field $\beta$ to be that of the standard magnetic flux of the supertube:
 \begin{equation}
\beta ~=~  \frac{R_y \,a^2}{\sqrt{2}\,\Sigma}\,(\sin^2\theta\, d\varphi_1 - \cos^2\theta\,d\varphi_2)\,, \qquad \Theta^{(3)}  ~\equiv~ d\beta  \,.  \label{betadefn}
\end{equation}
We use the following frames on the four dimensional base, $\cB$, with metric (\ref{basemet}):
\begin{equation}
e_1 ~=~ \frac{\Sigma^{1/2} }{(r^2 + a^2)^{1/2}} \, dr\,, \quad   e_2 ~=~\Sigma^{1/2}   \, d\theta\,, \quad   e_3 ~=~(r^2 + a^2)^{1/2}  \sin \theta  \, d\varphi_1\,, \quad   e_4 ~=~ r  \cos  \theta \, d\varphi_2 \,,
\label{frames1}
\end{equation} 
and introduce a standard basis for the self-dual two forms:
\begin{equation}\label{selfdualbasis}
\begin{aligned}
\Omega^{(1)} &\equiv \frac{dr\wedge d\theta}{(r^2+a^2)\cos\theta} + \frac{r\sin\theta}{\Sigma} d\varphi_1\wedge d\varphi_2 ~=~  \frac{1}{\Sigma \, (r^2+a^2)^\frac{1}{2}  \cos\theta} \,(e_1 \wedge e_2 +  e_3 \wedge e_4)\,,\\
\Omega^{(2)} &\equiv  \frac{r}{r^2+a^2} dr\wedge d\varphi_2 + \tan\theta\, d\theta\wedge d\varphi_1  ~=~  \frac{1}{\Sigma^\frac{1}{2}\, (r^2+a^2)^\frac{1}{2} \cos\theta} \,(e_1 \wedge e_4 +  e_2 \wedge e_3)   \,,\\
 \Omega^{(3)} &\equiv \frac{dr\wedge d\varphi_1}{r} - \cot\theta\, d\theta\wedge d\varphi_2~=~  \frac{1}{\Sigma^\frac{1}{2}\, r \sin\theta} \,(e_1 \wedge e_3 -  e_2 \wedge e_4)  \,.
\end{aligned}
\end{equation}
One may then write:
\begin{equation} 
\Theta^{(3)}  ~=~ d\beta 
~=~\frac{\sqrt{2}\,  R_y \,a^2 }{\Sigma^2} \, ( (r^2 + a^2)\cos^2 \theta \,  \Omega^{(2)}   - r^2 \sin^2 \theta \,  \Omega^{(3)} ) \,.
\label{Theta3form1}
\end{equation} 
In particular, $\Theta^{(3)}$ is self-dual.  

The first part of the solution is  defined by three more potential functions, $Z_I$, and magnetic $2$-forms, $\Theta^{(I)}$, $I=1,2,4$, that are required to satisfy the ``first layer'' of the linear system of equations governing all supersymmetric solutions of this theory:
 \begin{equation}\label{eqZ1Theta2}
 *_4 \mathcal{D} (\partial_v {Z}_1) =  \mathcal{D} \Theta^{(2)} \,,\quad \mathcal{D}*_4\mathcal{D}Z_1 = -\Theta^{(2)} \wedge d\beta\,,\quad \Theta^{(2)} =*_4 \Theta^{(2)} \,,
 \end{equation}
  \begin{equation}\label{eqZ2Theta1}
 *_4 \mathcal{D} ( \partial_v {Z}_2 ) =  \mathcal{D} \Theta^{(1)} \,,\quad \mathcal{D}*_4\mathcal{D}Z_2 = -\Theta^{(1)} \wedge d\beta\,,\quad \Theta^{(1)} =*_4 \Theta^{(1)} \,,
  \end{equation}
  \begin{equation}\label{eqZ4Theta4}
 *_4 \mathcal{D} (\partial_v {Z}_4) =  \mathcal{D}  \Theta^{(4)} \,,\quad \mathcal{D}*_4\mathcal{D}Z_4 = - \Theta^{(4)}\wedge d\beta\,,\quad  \Theta^{(4)}=*_4  \Theta^{(4)}\,.
\end{equation}
The operator, ${\mathcal D}$, acting on a $p$-form with legs on the four-dimensional base (and possibly depending on $v$), is defined by:
\begin{equation} 
{\mathcal D} \Phi ~\equiv~ d_{(4)} \Phi ~-~ \beta \wedge {\partial_v{\Phi}} \,, \label{Ddefn}
\end{equation} 
where $d_{(4)}$ denotes the exterior derivative on $\cB$.
The warp factor $\cP$ in  (\ref{sixmet}) is then determined by a quadratic form in the electric potentials:
\begin{equation}
\cP ~=~   Z_1\,Z_2 ~-~  Z_4^2\,. \label{Pform0}
\end{equation}

The remaining metric quantities are determined by the ``second layer'' of BPS equations:
\begin{equation}
 \begin{aligned}
D \omega + *_4 D\omega + \mathcal{F} \,d\beta 
~=~ & Z_1 \Theta^{(1)} +  Z_2 \Theta^{(2)} -   2\,Z_4 \Theta^{(4)} \,,  \\ 
 *_4D*_4\!\Bigl((\partial_v {\omega}) -\coeff{1}{2}\,D\mathcal{F}\Bigr) 
~=~& \partial_v^2 (Z_1 Z_2 - {Z}_4^2)  -((\partial_v {Z}_1)(\partial_v{Z}_2)  -(\partial_v {Z}_4)^2 )
\\ &   \qquad  -\coeff{1}{2} *_4\! \big(\Theta^{(1)}\wedge \Theta^{(2)} ~-~  \Theta^{(4)} \wedge  \Theta^{(4)} \big)\,.
\end{aligned}
\label{eqFomega}
\end{equation} 
We will study solutions to the BPS equations with mode dependence of the form:
\begin{equation}
\label{SSmodes1}
\chi_{k,m,n} ~\equiv~  \coeff{\sqrt{2}}{ R_y}\, (m+n) \, v ~+~  (k-m) \, \varphi_1 ~-~   m \,  \varphi_2    \,.
\end{equation}
We also define:
\begin{equation}\label{Deltadefn}
\Delta_{k,m,n}~\equiv~ \frac{a^k \, r^n }{(r^2+a^2)^{\frac{k+n}{2} }}\,\sin^{k-m}\theta\,\cos^m\theta\,.
\end{equation}
The smoothness of the solutions requires $k$ to be a positive integer and  $m$, $n$ to be non-negative integers with  $m\le k$. This restriction has a clear holographic interpretation in the description of the dual CFT states~\cite{Bena:2015bea,Bena:2016agb,Bena:2016ypk}. 

The most general solution to the first layer is known for the single-bubble solutions that are built upon the circular supertube \cite{Niehoff:2013kia,Shigemori:2013lta,Giusto:2013bda,Bena:2015bea,Bena:2016ypk}.  This family of solutions can be represented by a  superposition of the following single-mode solutions for the pair $(Z_4, \Theta^{(4)})$:
\begin{equation}
\begin{aligned}
Z_4  ~=~ &   b_4 \frac{ R_y}{\Sigma} \;\! \Delta_{k,m,n}   \;\! \cos \chi_{k,m,n} \,, \\
\Theta^{(4)}   ~=~ & - {\sqrt{2}} \,b_4 \;\!  \Delta_{k,m,n} \, \Big[\Big((m+n) \, r\, \sin \theta + n\, \Big(\frac{m}{k} -1\Big)\,  \frac{\Sigma}{r \, \sin \theta} \Big)\, \sin \chi_{k,m,n} \,\Omega^{(1)} \\
&\qquad \qquad \qquad\qquad ~+~ \cos  \chi_{k,m,n} \, \Big(m\, \Big(\frac{n}{k} + 1\Big)\, \Omega^{(2)} ~+~  n\,\Big(\frac{m}{k} - 1\Big)\, \Omega^{(3)}\Big)\Big] \,,
\end{aligned}
\label{Z4Theta4form}
\end{equation}
with similar expressions for $(Z_1, \Theta^{(2)})$ and $(Z_2, \Theta^{(1)})$, with a priori independent coefficients.

The general families of regular solutions for the second layer have not been classified.  Classes of single-mode solutions are known \cite{Bena:2015bea, Bena:2016agb, Bena:2016ypk, Bena:2017geu} and some multi-mode solutions have been obtained \cite{Bena:2015bea}.  Here we will focus entirely on the families of single-mode solutions obtained in~\cite{Bena:2016ypk} and further studied in~\cite{Bena:2017geu}. 
 
\subsection{Coiffured single-mode solutions}
\label{ss:SingModeSols}

In the maximally-rotating supertube solution, the data of the first layer of BPS equations takes the following simple form:
\begin{equation}
Z_1  ~=~ \frac{Q_1}{\Sigma} \,, \qquad Z_2  ~=~ \frac{Q_5}{\Sigma}  \,,  \qquad Z_4  ~=~ 0\,, \qquad\qquad\Theta^{(I)}  ~=~ 0\,, \quad I = 1,2,4  \,.
\label{stsol}
\end{equation}
To this solution we add a single fluctuating mode  by taking $Z_4$, $\Theta_4$ to be given by (\ref{Z4Theta4form}) and by taking:
\\
\begin{equation}
\begin{aligned}
Z_1  ~=~ & \frac{Q_1}{\Sigma} \left( 1 + \frac{b_4^2}{2a^2+b^2} \;\! \Delta_{2k,2m,2n}  \;\! \cos \chi_{2k,2m,2n} \right) \,,
\qquad Z_2  ~=~ \frac{Q_5}{\Sigma}  \,, \qquad\Theta^{(1)}  ~=~ 0 \,,  \\
\Theta^{(2)}  ~=~ & {}-  b_4^2 \;\!\frac{ R_y}{{\sqrt{2}}\, Q_5}  \;\!  \Delta_{2k,2m,2n} \, \Big[\Big(2(m+n) \, r\, \sin \theta + 2n\, \Big(\frac{m}{k} -1\Big)\,  \frac{\Sigma}{r \, \sin \theta} \Big)\, \sin \chi_{2k,2m,2n} \,\Omega^{(1)} \\
&\qquad \qquad \qquad\qquad \quad~+~ \cos  \chi_{2k,2m,2n} \, \Big(2m\, \Big(\frac{n}{k} + 1\Big)\, \Omega^{(2)} ~+~  2n\,\Big(\frac{m}{k} - 1\Big)\, \Omega^{(3)}\Big)\Big]  \,. \cr
\end{aligned}
\label{ZThetaform1}
\end{equation}
Observe that the Fourier frequencies appearing in $(Z_1,\Theta^{(2)})$ are twice those appearing in $(Z_4,\Theta^{(4)})$ and that the Fourier coefficients of these modes have been tuned in terms of the square of the Fourier coefficients of $(Z_4,\Theta^{(4)})$.  This is an example of the procedure known as ``coiffuring'' \cite{Mathur:2013nja,Bena:2013ora}. The problem is that generic fluctuations for the solutions to the first layer of BPS equations typically lead to singular solutions in the second layer.  This may be related to the non-linear instabilities that have been suggested in  \cite{Eperon:2016cdd,Keir:2016azt,Marolf:2016nwu,Eperon:2017bwq}; this is currently under investigation.  Coiffuring solves this problem by tuning other excitations to remove the singularities in the solutions to the second layer of BPS equations.  For a single mode the result is particularly simple:  All dependence on $(v, \varphi_1,\varphi_2)$ cancels in the sources for the second layer of BPS equations and in the warp factor, $\cP$.  As a result, for a single mode, the entire metric (\ref{sixmet}) is independent of  $(v, \varphi_1,\varphi_2)$.  All that remains of the fluctuations is the ``RMS values'' proportional to $b_4^2$.  

The warp factor $\cP$ now reduces to:
\begin{equation}
\cP 
~\,=\,~ \frac{Q_1 Q_5}{\Sigma^2}\left( 1 - \frac{b_4^2}{2a^2+b^2} \Delta_{2k,2m,2n} \right) \,. \label{Pform1}
\end{equation}
It was shown in \cite{Bena:2016ypk} that this is positive definite for all $r$ and $\theta$.

Next, one must solve the second layer of BPS equations.  Since the coiffured sources are independent of $(v, \varphi_1,\varphi_2)$, this means that we can use the following Ansatz:
\begin{equation}
\omega ~\equiv~ \omega_1 \, d\varphi_1 + \omega_2 \, d\varphi_2 ~=~ \omega_0 + \hat  \omega_1 (r,\theta) \,  d\varphi_1 +  \hat \omega_2 (r,\theta) \,  d\varphi_2 \,, \qquad \cF ~=~  \cF(r,\theta)  \,, \label{Fomforms}
\end{equation}
where $\omega_0$ the angular momentum vector of the round supertube: 
\begin{equation} 
\omega_0 ~ \equiv~ \frac{R_y\, a^2}{\sqrt{2} \, \Sigma} \, ( \sin^2 \theta \, d\varphi_1 +   \cos^2 \theta \, d \varphi_2 ) \,,
\label{om0}
\end{equation} 
and where $\hat \omega_1, \hat\omega_2$ and $\cF$ are determined by solving (\ref{eqFomega}), with the sources on the right-hand side given by the fluctuating solution to the first layer of BPS equations described above.

The general family of $(k, m, n)$ single-mode superstratum solution was obtained in \cite{Bena:2016ypk}.  Regularity at $r=0,\theta=\pi/2$ (the ``supertube regularity'' condition) imposes the constraint: 
\begin{equation}
\frac{Q_1Q_5}{R_y^2}  ~=~   a^2 +  \frac{b^2}{2}  \,,  \qquad\qquad 
b^2 ~\equiv~ \bigg[ {k \choose m}{k+n-1 \choose n} \bigg]^{-1}b_4^2 \,.
  \label{regularity-b-b4}
\end{equation}
The explicit solutions for $(k, m, n)=(1,0,n)$ and $(k, m, n)=(2, 1, n)$  were studied in detail in  \cite{Bena:2016ypk}  and  \cite{Bena:2017geu}  respectively, and we will exhibit them momentarily.

In  \cite{Bena:2016ypk} it was shown that the complete $(k, m, n)$ solution has the following values of the conserved five-dimensional angular momenta $j$, $\tilde{j}$, and $y$-momentum $\np$: 
\begin{equation}
  j =  \frac{\mathcal{N}}{2} \left({a^2} + \frac{m}{k} \, b^2\right), \hspace*{.5cm} \tilde{j} = \frac{\mathcal{N}}{2} a^2,  \hspace*{.5cm}  \np  = \frac{\mathcal{N}}{2} \frac{m+n}{k} \, b^2\,,   \label{conschgs}
\end{equation}
where $ \mathcal{N} \equiv n_1 n_5 R_y^2 / (Q_1 Q_5)$, and $n_1, n_5$  are the numbers of D1 and D5 branes.   It was proposed in \cite{Bena:2016ypk} that these solutions  
are holographic duals of coherent superpositions of CFT states of the form:
 \begin{equation}
(|\!+\!+\rangle_1)^{N_1 }
\biggl(\frac{(J^+_{-1})^{m}}{m!} \frac{(L_{-1}- J^3_{-1})^n}{n!} |00\rangle_k\biggr)^{N_{k,m,n} }\,,  \label{eq:dualstates}
\end{equation}
for all values of $N_1$ such that $N_1+  k N_{k,m,n} = N$. For an explanation of the above notation, see~\cite{Bena:2016ypk}. The values of the conserved  charges imply that, within the coherent superposition, the average numbers of $|\!+\!+\rangle_1$ and $|00\rangle_k$ strands are given by $\mathcal{N}a^2$ and  $\mathcal{N}b^2/(2k)$ respectively.

\section{The special families of superstrata metrics} 
\label{sec:Special-properties}

We now examine the details of the solutions for which the parameters $(k, m,n)$ take the values $(1,0,n)$, $(2,1,n)$, and $(2,0,n)$.    The solution for $(k\! =\! 2, m\! =\! 0)$ is included simply to illustrate that properties (i) and (ii) do not hold in general, since neither property holds for this solution.  We will therefore not discuss the details of this particular solution beyond writing down the metric, and we will focus on the other two families.

For $k\!=\!1$, $m\!=\!0$ and general $n > 0$, the solution to the second layer of BPS equations is  \cite{Bena:2016ypk}: 
\begin{equation}
\cF ~=~  - \frac{b_4^2}{a^2}\,\Big(1 - \frac{r^{2n}}{(r^2+a^2)^{n}}\Big) \,, \qquad \omega ~=~ \omega_0 ~+~   \frac{R_y\, b_4^2}{\sqrt{2}\,  \Sigma}\, 
 \Big(1 - \frac{r^{2n}}{(r^2+a^2)^{n}} \Big) \, \sin^2\theta\, d\varphi_1\,. 
\label{sol10n}
\end{equation}
For $k\!=\!2$, $m\!=\!1$ and general $n > 0$, we have \cite{Bena:2017geu}\footnote{To arrive at the following expression we have taken the results from Section 6.1 of  \cite{Bena:2017geu} and undone the gauge transformations described in Section 2.2 of that paper.  We have thus ensured that (\ref{sol10n}) and (\ref{sol21n}) are solutions in the conventions of this paper. }:
\begin{equation}
\begin{aligned}
\cF ~=~ & - \frac{b^2 }{a^2} + 
 \frac{b_4^2 \, r^{2n}}{4 \,(r^2 +a^2)^{n+2}}\,    \left( \Sigma 
+ \frac{2\, r^2\, (r^2 +a^2) }{(n+1)\,a^2}\right) \,, \\ 
\omega_1 ~=~ & \frac{ R_y}{\sqrt{2}\,\Sigma}\bigg[ \,(a^2 + b^2) \sin^2 \theta
-\frac{b_4^2}{2} \,  \frac{  r^{2n}\,\sin^2 \theta}{(r^2 +a^2)^{n+1}}\,
\bigg(  \frac{r^2}{2\, (n+1)} +a^2\, \cos^2 \theta  \bigg) \bigg]  \,, \\ 
\omega_2 ~=~ &\frac{ R_y}{\sqrt{2}\,\Sigma}\,\bigg[ \, a^2 \cos^2 \theta ~+~ \frac{b_4^2}{2}  \,\frac{r^{2(n+1)} \,\cos^2 \theta}{(r^2 +a^2)^{n+2}}\, 
\bigg(\frac{(r^2 +a^2)}{2\,(n+1)}   +a^2\, \sin^2 \theta \bigg)\,\bigg]\,. 
\end{aligned}
\label{sol21n}
\end{equation}
\vspace{-1mm}

\noindent
Finally, for $k\!=\!2$, $m\!=\!0$ the solution to the second layer of BPS equations is:
\begin{equation}
\begin{aligned}
\cF ~=~ & - \frac{b_4^2}{(n+1)^2\, a^4} \bigg[  n a^2 - r^2\bigg( 1 - \frac{r^{2n}}{(r^2 +a^2)^{n}} \bigg) \\
& \qquad~+~ \bigg( \bigg( 1 - \frac{r^{2n}}{(r^2 +a^2)^{n}} \bigg)(2r^2 + (2n+1) a^2)  - 2 n a^2  - \frac{n^2 a^4 \,r^{2n}}{(r^2 +a^2)^{n+1}} \bigg) 
\sin^2 \theta \bigg]  \,, \\ 
\omega_1 ~=~ & \frac{ R_y}{\sqrt{2}\,\Sigma}\bigg\{  \, a^2 \sin^2 \theta + \frac{b_4^2}{(n+1)^2}\, \bigg[ (n+1) \bigg(  1 - \frac{r^{2n}}{(r^2 +a^2)^{n}}  -\frac{ n \, a^2 \, r^{2n}}{(r^2 +a^2)^{n+1}} \bigg) \, \sin^2 \theta \\
& \qquad  \qquad~-~ \bigg( \frac{r^2}{a^2}\,\bigg( 1 - \frac{r^{2n}}{(r^2 +a^2)^{n}} \bigg) ~-~ n\bigg) \, \cos^2 \theta \bigg]  \, \sin^2 \theta \, \bigg\},\\
\omega_2 ~=~ &\frac{ R_y}{\sqrt{2}\,\Sigma}\,\bigg[ \, a^2 \cos^2 \theta ~-~ \frac{b_4^2}{(n+1)^2}\, \bigg[ \frac{r^2}{a^2}\,\bigg( 1 - \frac{r^{2n}}{(r^2 +a^2)^{n}} \bigg) -  \frac{n \, r^{2n+2}}{(r^2 +a^2)^{n+1}} \bigg]\, \sin^2 \theta\,  \cos^2 \theta  \,.
\end{aligned}
\label{sol20n}
\end{equation}
\vspace{-1mm}

\noindent
Regularity requires that $b$ and $b_4$ are related via the general relation (\ref{regularity-b-b4}), which evaluates to:
\begin{equation}
 b^2 ~=~ b_4^2  \quad{\rm for } \quad k =1 \,\qquad \qquad {\rm and} \qquad\qquad b^2 ~=~ \frac{b_4^2}{2(n+1)} \quad{\rm for } \quad k=2 \,.
  \label{eq:regularity-b-b4-21n}
\end{equation}
Note that killing off all the non-trivial modes by setting $b_4 = b =0$ in (\ref{sol10n})--(\ref{sol20n}) reduces these expressions to $\omega_0$ in (\ref{om0}).

\subsection{The $(1,0,n)$ family of solutions}
\label{ss:Example1}

\subsubsectionmod{Metric fibration}
\label{ss:Example1-metric}

The value of $\cP$ in the starting supertube solution is $Q_1 Q_5/\Sigma^2$. It is convenient to factor this off and introduce the quantity: 
\begin{equation}
\Lambda ~\equiv~ \frac{\sqrt{\cP}\,\Sigma}{\sqrt{Q_1 Q_5}} ~=~ \sqrt{ 1 - \frac{a^2\,b^2}{(2 a^2 +b^2)} \, \frac{r^{2n}}{(r^2 +a^2)^{n+1}} \, \sin^2 \theta  } \,.
  \label{Lambdadef1}
\end{equation}

\noindent
The six-dimensional metric, (\ref{sixmet}), can then be re-written as: 
\bea
 ds_6^2 &=& \sqrt{Q_1 Q_5} \,  \frac{\Lambda}{F_2(r)} \bigg[ \frac{F_2(r) \, dr^2}{r^2 + a^2}  \,-\, \frac{2\,F_1(r)}{a^2 (2 a^2 + b^2)^2 \;\! R_y^2 }\bigg(dv + \frac{a^2\,(a^4 + (2 a^2 +b^2) r^2)}{F_1(r)} du \bigg)^2 \cr
& & \qquad \quad \qquad \qquad~ +\, \frac{2 \, a^2 \,r^2 \,(r^2 + a^2)\, F_2(r)}{F_1(r) \, R_y^2} \, du^2 \bigg] \label{Redform1a} \\
&& ~+\sqrt{Q_1 Q_5} \, \bigg[  \Lambda \, d\theta^2  \,+\,  
\frac{1}{\Lambda} \sin^2 \theta \, \Big(d\varphi_1 -  \frac{a^2}{(2a^2 +b^2)}\frac{\sqrt{2}}{R_y} (du + dv) \Big)^2  \cr
&& \qquad\qquad\quad~~ + \, \frac{F_2(r)}{\Lambda} \cos^2 \theta \, \Big(d\varphi_2 + \, \frac{1}{(2a^2 +b^2)\, F_2(r)}\frac{\sqrt{2}}{R_y}  \left[ a^2 (du - dv) -  b^2 \, F_0 (r) dv \right] \Big)^2  \bigg] \,,
\nonumber
\eea
where the functions, $F_i(r)$, are defined by:
\begin{equation}
\begin{aligned}
F_0(r) ~\equiv~ & 1 - \frac{r^{2n}}{(r^2 +a^2)^{n}}  \,, \qquad F_1(r) ~\equiv~ a^6 - b^2 \, (2 a^2 + b^2) \,r^2 \, F_0(r) \,, \\
F_2(r) ~\equiv~ &  1  - \frac{a^2\, b^2}{(2 a^2 + b^2) } \,\frac{ r^{2n}}{(r^2 +a^2)^{n+1}}   \,.
\end{aligned}
  \label{Fdefs}
\end{equation}

From (\ref{Redform1a}) it is elementary to evaluate the determinant of the internal metric along $\cS$. Recalling that (\ref{Omdefn}) defines the warp factor required by dimensional reduction, we find:
\begin{equation}
\Omega^{-2} ~=~   \frac{\Lambda}{ F_2(r)}  \,.
  \label{Omega1}
\end{equation}
In  (\ref{Redform1a}) we have extracted this warp factor from the first part of the metric and so the metric terms inside the first set of square brackets yield the metric, $\hat g_{\mu \nu}$, on $\cK$ defined in (\ref{fiberform}).  The resulting three-dimensional metric is, indeed, purely a function of $r$,
\begin{equation}
\begin{aligned}
\hat g_{\mu \nu}\, dx^\mu dx^\nu
     ~=~& \sqrt{Q_1 Q_5} \;\! \Bigg[ 
		\frac{F_2(r) \, dr^2}{r^2 + a^2} 
		+\, \frac{2 \, a^2 \,r^2 \,(r^2 + a^2)\, F_2(r)}{F_1(r) \, R_y^2} \, du^2 \\
 & \qquad \qquad \qquad \,-\, \frac{2\,F_1(r)}{a^2 (2 a^2 + b^2)^2 \;\! R_y^2 }\bigg(dv + \frac{a^2\,(a^4 + (2 a^2 +b^2) r^2)}{F_1(r)} du \bigg)^2 \;\! \Bigg]  \;\!.
\end{aligned}
\label{dssq12}
\end{equation}

We define the following one-forms on the three-dimensional base, $\cK$:
\begin{equation}
A^{(1)} ~\equiv~   - \frac{a^2}{(2a^2 +b^2)} \frac{\sqrt{2}}{R_y}(du + dv)  \,, \qquad  A^{(2)} ~\equiv~ \frac{1}{(2a^2 +b^2)\, F_2(r)}  \frac{\sqrt{2}}{R_y}  (a^2 (du - dv) - b^2 \, F_0 (r) dv)\,.
  \label{Max1}
\end{equation}
We now observe that the off-diagonal components, ${B^i}_\mu$  of the fibration form of the metric (\ref{fiberform}) can be written as
\begin{equation}
{B^i}_\mu dx^\mu ~=~{K_{(1)}^{\phantom{a}}}^{\!\!\! i} \; A^{(1)} +   {K_{(2)}^{\phantom{a}}}^{\!\!\! i} \; A^{(2)}   \,,
  \label{KKvecs1}
\end{equation}
where ${K_{(1)}}^{\!\!\! M}  = (0,0,0,0,1,0)$ and ${K_{(2)}}^{\!\!\! M} = (0,0,0,0,0,1)$ are the components of the Killing vectors $\partial/\partial\varphi_1$ and $\partial/\partial\varphi_2$ respectively.   This means that, under dimensional reduction, the vector fields, $A^{(1)}$  and $A^{(2)}$ are massless electromagnetic potentials on $\cK$.    Thus, not only is the metric $\hat g_{\mu \nu}$ independent of the coordinates on $\bbS^3$,  but so are the dynamical components of the metric on $\cS$.  In light of Property (iii) and  the comments made at the end of Section \ref{sec:DimRedGen} about the absence of scalar excitations and shape modes, the dynamics of the six-dimensional {\it metric excitations} in this solution reduces to dynamics of the metric and massless vector fields on $\cK$.  Of course, one should recall that in the complete  six-dimensional solution, the three-form fields depend upon $v, \varphi_1$ and $\varphi_2$, and so do not reduce to massless fields in three dimensions. It is, however, still possible that such tensor gauge modes give rise, in a consistent truncation, to a collection of massless and massive fields on $\cK$. This possibility is currently under investigation. 

\subsubsectionmod{Geodesics}
\label{ss:Example1-geodesics}

Since the six-dimensional metric is independent of $(u,v,\varphi_1, \varphi_2)$, this means that the corresponding momenta are conserved:
\begin{equation}
L_1 ~=~ {K_{(1) M }} \frac{dz^M}{d \lambda} \,, \qquad L_2 ~=~ {K_{(2)  M }}  \frac{dz^M}{d \lambda} \,,\qquad  P ~=~ {K_{(3)   M }}  \frac{dz^M}{d \lambda} \,, \qquad E ~=~ {K_{(4)   M }}  \frac{dz^M}{d \lambda}   \,,
  \label{ConsMom}
\end{equation}
where the $K_{(I)}$  are the Killing vectors: $K_{(1)}  = \frac{\partial}{\partial \varphi_1}$, $K_{(2)}  = \frac{\partial}{\partial \varphi_2}$, $K_{(3)}  = \frac{\partial}{\partial v }$ and $K_{(4)}  = \frac{\partial}{\partial u}$.   In addition, there is the standard quadratic conserved quantity coming from the metric: 
\begin{equation}
\varepsilon ~\equiv~ g_{MN} \, \frac{dz^M}{d \lambda} \, \frac{dz^N}{d \lambda}  \,.
  \label{MetInt}
\end{equation}

These conservation laws determine all the velocities except $v_r \equiv \frac{d r}{d \lambda}$ and $v_\theta \equiv \frac{d\theta}{d \lambda}$.  In principle 
(\ref{MetInt}) allows exchange of energy between $v_r$ and $v_\theta$, and this could generate interesting trapping of geodesics: as a particle falls in, some of its $v_r$ is traded for $v_\theta$ and thus the particle may lose radial momentum and be prevented from returning to where it started. 

However, at least for null geodesics in the $(1,0,n)$ family of metrics, there is a hidden symmetry: there is an additional conserved quantity, that is quadratic in momenta. The additional conserved quantity can be found by separating variables in the massless Hamilton-Jacobi equation, and takes the form: 
\begin{equation}
\Xi ~\equiv~ \xi_{MN} \, \frac{dz^M}{d \lambda} \, \frac{dz^N}{d \lambda} ~\equiv~Q_1 Q_5 \, \Lambda^2 \, v_\theta^2 ~+~ \frac{L_1^2}{\sin^2 \theta}~+~ \frac{L_2^2}{\cos^2 \theta} \,.
  \label{ConKtens}
\end{equation}
One can verify that for any geodesic one has 
\begin{equation}
 \frac{d}{d \lambda}\Xi ~=~R_y\, v_\theta \, \bigg(\frac{\partial \Lambda}{\partial \theta}\bigg) \, \bigg( g_{MN} \, \frac{dz^M}{d \lambda} \, \frac{dz^N}{d \lambda}\bigg) \,,
  \label{ConKtens2}
\end{equation}
which vanishes on null geodesics.

\begin{figure}
\centerline{\includegraphics[width=4.4in]{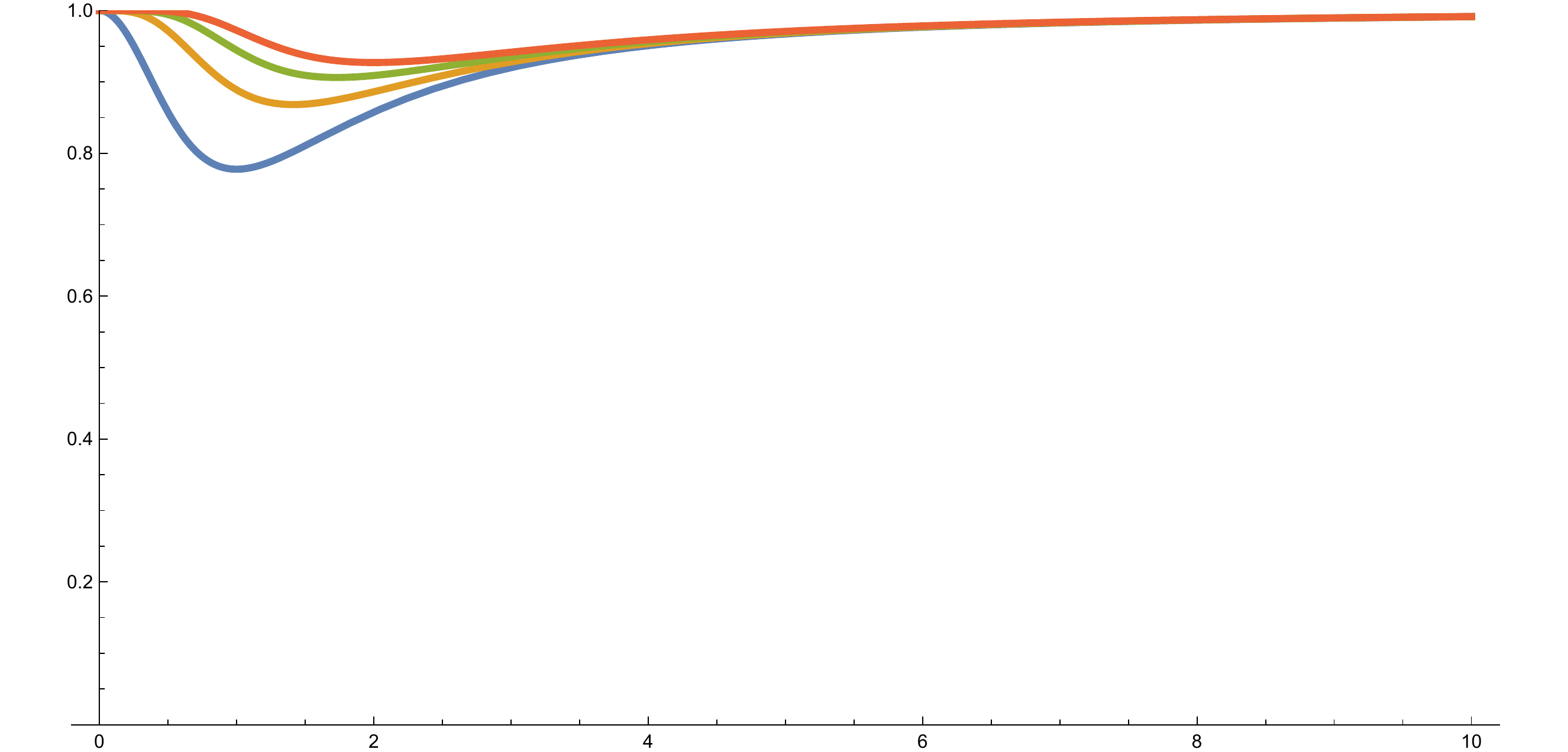}}
\setlength{\unitlength}{0.1\columnwidth}
\caption{\it 
Plot of $F_2(r)$ for $a=1$, $b=4$ and $n=1,2,3,4$.  The curve with the largest dip has $n=1$ and, as (\ref{F2min}) implies, the size of the dip decreases with $n$. }
\label{fig:F2vars}
\end{figure}

Were it not for the presence of $\Lambda^2$ in   (\ref{ConKtens}), this conserved quantity would be the total angular momentum on the round $\bbS^3$, and the motion on $\cS$ would be essentially decoupled from that on $\cK$.  However, because of the the factor of $\Lambda^2$, these motions are not decoupled.  On the other hand, the factor of $\Lambda^2$ only exerts a minor influence on geodesic motion. To see this, first observe that 
\begin{equation}
\Lambda^2|_{r=0} ~=~ 1 \,, \qquad \Lambda^2 ~\to~ 1 \quad {\rm as}\quad r \to \infty\,, \qquad\Lambda^2|_{\theta=\frac{\pi}{2}} ~=~ F_2(r) \,.
  \label{limits}
\end{equation}
In fact $\Lambda^2$ is very close to $1$ for most values of $(r,\theta)$, and its maximum deviation from $1$ is at $\theta=\frac{\pi}{2}$ where it is given by $F_2(r)$.  The function $F_2(r)$ is minimized at $r = a \sqrt{n}$, and has a minimum value of 
\begin{equation}
1  ~-~ \frac{b^2}{2 a^2+ b^2} \,\frac{n^n}{(n+1)^{n+1}} ~>~ \frac{3}{4} \,.
  \label{F2min}
\end{equation}
Moreover, as one can see in Fig.~\ref{fig:F2vars}, the variation from $1$ takes place in a  short interval around $r = a \sqrt{n}$.  The region around $r = a \sqrt{n}$ is also the region where the microstate structure, in the form of momentum-carrying waves, is concentrated. 

This means that $v_\theta$ will increase briefly as it passes through $r = a \sqrt{n}$ and this can, in turn, change the value of $\theta$ and the value of $v_r$.  However, this effect is quite localized and makes little difference to the asymptotic values or $v_r$ and $r$.  A free particle falling in from outside the throat will bounce off the center and escape the throat.  The only geodesics that are affected significantly by the microstate structure are the ones that are already localized near $r = a \sqrt{n}$.

\subsubsectionmod{The wave equation}
\label{ss:Example1-waves}

Consider the six-dimensional scalar wave equation 
\begin{equation}
\frac{1}{\sqrt{- \det{(g_{MN})}} }\, \frac{ \partial}{\partial z_P} \,\bigg( \sqrt{- \det{(g_{MN})}}\, g^{PQ} \frac{ \partial}{\partial z_Q} \Phi \, \bigg) ~=~ \frac{M^2}{ \sqrt{Q_1 Q_5}\, \Lambda} \, \Phi \,,
  \label{WaveOp}
\end{equation}
where the factor of $\Lambda^{-1}$ has been included in the ``mass term'' on the right-hand side for reasons that will become apparent below.

Consider a generic mode for $\Phi$ of the form  
\begin{equation}
\Phi ~=~ K(r) \, S(\theta) \, e^{i \big(\frac{\sqrt{2}}{R_y}\omega\;\! u  + \frac{\sqrt{2}}{R_y} p \;\! v + q_1\;\! \varphi_1 + q_2\;\! \varphi_2 \big)}\,.
  \label{SepFns}
\end{equation}
One then finds that the wave equation separates, yielding: 
\bea
 \frac{1}{r} \partial_r \Big(  r(r^2 +a^2) \,  \partial_r K \Big)\,+
\left( \frac{a^2 (\omega+p +q_1)^2}{r^2 + a^2}
-\frac{a^2 (\omega-p-q_2)^2}{r^2}
\right) K \quad &&  \label{SepEqns1}\\
 +\,\, \frac{b^2 \omega  \left(2 a^2 p+ F_0(r) \! \left[2 a^2 (\omega + q_1) + b^2 \omega \right]\right)}{a^2 \left(r^2 + a^2 \right)} K
  &=& (\lambda+M^2) \, K \,, 
	\nonumber \\
\frac{1}{\sin \theta \cos \theta} \partial_\theta \big( \sin \theta \cos \theta \,  \partial_\theta S \big)\,-\, \left(\frac{q_1^2}{\sin^2 \theta} \,+\, \frac{q_2^2}{\cos^2 \theta}\right)\,S     &=& - \lambda \, S \,,
  \label{SepEqns2}
\eea
for some eigenvalue $\lambda$.  

Observe that the second equation (\ref{SepEqns2}) is the eigenvalue problem for the Laplace operator on the {\it round} $\,\bbS^3$.  The regular modes are therefore given in terms of Jacobi polynomials: 
\begin{equation}
S(\theta) ~=~ \sin^{q_1} \theta \,\cos^{q_2}  \theta\, P_\bn^{(q_1, q_2)}(\cos 2\theta) \,, 
  \label{SasJP}
\end{equation}
where
\begin{equation}
P_\bn^{(q_1, q_2)}(x) ~=~  \frac{(q_1 +1)_\bn}{\bn !} \, {}_2 F_{1}\big(-\bn ,1+\coeff{1}{2}(\ell+q_1+q_2)  ; q_1+1 ;\coeff{1}{2}(1-x)\big) \,, 
  \label{JacPolys}
\end{equation}
and $(y)_\bn  = \prod_{m=0}^\bn (y-m)$ is Pochhammer's symbol.  The quantum numbers $(\ell,\bn)$ are defined by
\begin{equation}
 \lambda ~=~  \ell(\ell+2)\,, \qquad \bn~=~\coeff{1}{2}(\ell-q_1-q_2) \,.
  \label{intparams}
\end{equation}
For the modes (\ref{SepFns}) to be single-valued and regular, and for $P_\bn^{(q_1, q_2)}$ to be a polynomial, one must have 
\begin{equation}
 q_1, q_2, \ell, \bn  ~\in~  \ZZ \,, \qquad  q_1, q_2 ~\ge~ 0 \,, \qquad \ell ~\equiv~q_1+q_2 \ {\rm mod}\   2 \,.
  \label{intconds}
\end{equation}

The radial equation is considerably more involved, because of the presence of terms proportional to $b^2$, which encode the massless scattering from the detailed structure of the superstratum. Note that setting $r = a \sinh \xi$ turns the differential operator part of (\ref{SepEqns1}) to a more canonical form:
\begin{equation}
\frac{1}{\sinh \xi \cosh \xi} \partial_\xi \big( \sinh \xi \cosh \xi \,  \partial_\xi  K \big) \,.
  \label{SepOp1}
\end{equation}
Note also that in the limit $b \to 0$, the background becomes the decoupling limit of the circular supertube solution, and the radial equation simply becomes the hypergeometric equation~\cite{Lunin:2001dt}. It has been known for some time that the massless scalar wave equation is separable in the circular supertube solution~\cite{Lunin:2001dt}, as well as in solutions obtained by spectral flow thereof~\cite{Giusto:2004ip,Giusto:2012yz,Chakrabarty:2015foa}, and related black hole solutions~\cite{Cvetic:1997uw}. 
The $(1,0,n)$ family of single-mode superstrata is considerably more complicated than these solutions, and so it is quite remarkable that separability is preserved.

Finally, note that the mass term in (\ref{WaveOp}) has descended to a mass term in three dimensions, as is evident in (\ref{SepEqns1}).  Indeed the total three-dimensional mass of the scalar field is the sum of the explicit mass, $M^2$, and the eigenvalue,  $\lambda$.

\subsection{The $(2,1,n)$ family of solutions}
\label{ss:Example2}

We now analyze the $(2,1,n)$ family of solutions. The warp factor, $\Lambda$, now takes the form:
\begin{equation}
\Lambda ~\,=~\,  \sqrt{ 1 -\frac{2 (n+1)  \,a^4\,b^2}{(2 a^2 +b^2)} \, \frac{r^{2n}}{(r^2 +a^2)^{n+2}} \, \sin^2 \theta\, \cos^2 \theta  } \ \,.
  \label{Lambdadef2}
\end{equation}
Note that, compared to the corresponding warp factor in the $(1,0,n)$ solutions, (\ref{Lambdadef1}),  the warp factor involves a higher harmonic mode with a stronger fall-off at infinity.  This means that the non-trivial profile in $\Lambda$ is even smaller than the profiles depicted in  Fig.~\ref{fig:F2vars}.   The profile is sharply peaked around 
\begin{equation}
r ~=~ \sqrt{\frac {n}{2}}\, a  \,.
  \label{peak2}
\end{equation}
We introduce the coordinates 
\bea
\psi &=& \varphi_1 + \varphi_2 \,,  \qquad \phi ~=~ \varphi_2 - \varphi_1 \,,
\eea
and the functions, $H_i(r)$:
\begin{equation}
\begin{aligned}
H_0(r) ~\equiv~ & 1 - \frac{r^{2n+2}}{(r^2 +a^2)^{n+1}}  \,, \qquad H_1(r) ~\equiv~1 + \frac{a^2\, b^2 }{2\,(2 a^2 + b^2) } \,\frac{ r^{2n}}{(r^2 +a^2)^{n+1}}\,, \\
H_2(r) ~\equiv~ &  1  - \frac{a^4\, b^2\,(n+1)}{2\,(2 a^2 + b^2) } \,\frac{ r^{2n}}{(r^2 +a^2)^{n+2}}    \,.
\end{aligned}
  \label{Hdefs}
\end{equation}
In terms of these, the six-dimensional metric, (\ref{sixmet}), can be re-written as: 
\begin{equation}
\begin{aligned}
 ds_6^2 ~=~ &  \sqrt{Q_1 Q_5} \,   \Lambda \;\! \bigg[ \;\! \frac{dr^2}{r^2 + a^2} \,+\, \frac{2 \, r^2 \,(r^2 + a^2) }{a^4 \,R_y^2} \, dv^2  \\ &\qquad \qquad\quad~   - \, \frac{2}{a^4\,(2 a^2 + b^2)^2\, R_y^2 \, H_2(r)}\Big(  a^4\,(du+ dv) +  (2 a^2 + b^2)\,r^2  \, H_1(r) \, dv \Big)^2  \;\!  \bigg]  \\
  & +\sqrt{Q_1 Q_5} \, \bigg[ \;\! \Lambda \, d\theta^2 \,+\, \frac{H_2(r)}{4\Lambda}
	\Big( d\psi + \hat A^{(\psi)}	\Big)^2 + \frac{H_2(r)}{4\Lambda}\cos 2\theta	\big( d\psi +\hat A^{(\psi)}	\big)	\big( d\phi + \hat A^{(\phi)}	\big) 
	\\
& \qquad \qquad \qquad\qquad \qquad+ \frac{\cos^2( 2\theta) H_2(r) + \sin^2(2\theta)}{4\Lambda}	\big( d\phi + \hat A^{(\phi)}	\big)^2 \;\! \bigg] 
\,,
\end{aligned}
\label{Redform2a}
\end{equation}
where the vector fields $\hat A^{(\psi)}$ and  $\hat A^{(\phi)}$ are given by:
\begin{equation}
\begin{aligned}
\hat A^{(\psi)} ~=~ & \; \frac{2\sqrt{2}}{{R_y}} \, \left[
-\frac{1}{2} dv 
+ \cos 2\theta \, \frac{1-H_2(r)}{H_2(r)} 
\left( \frac{a^2}{2a^2+b^2}(du+dv)+\frac{r^2}{a^2} H_1(r) dv
\right)
\right] \,, \\
\hat A^{(\phi)}~=~ & \; \frac{\sqrt{2}}{{R_y}}   \, \bigg[  
\frac{2a^2 du - b^2 H_0(r) dv}{(2 a^2 + b^2)}   \bigg]  \,.
\end{aligned}
  \label{Bdefs}
\end{equation}

As before,  it is elementary to evaluate the determinant of the internal metric along $\cS$ and use (\ref{Omdefn}) to obtain the warp factor required by dimensional reduction:
\begin{equation}
\Omega^{-2} ~=~   \frac{\Lambda}{ H_2(r)}  \,.
  \label{Omega2}
\end{equation}
The dimensionally-reduced metric on $\cK$, defined in (\ref{fiberform}), is then:
\begin{equation}
\begin{aligned}
 ds_{1,2}^2 ~=~  &   \sqrt{Q_1 Q_5}  \,H_2(r)   \;\! \bigg[ \frac{dr^2}{r^2 + a^2} \,+\, \frac{2 \, r^2 \,(r^2 + a^2) }{a^4 \,R_y^2} \, dv^2  \\ 
 &\qquad\qquad~~-\, \frac{2}{a^4 \,(2 a^2 + b^2)^2\, R_y^2 \, H_2(r)}\Big(  a^4\,(du+ dv) +  (2 a^2 + b^2)\,r^2  \, H_1(r) \, dv \Big)^2   \bigg]    \,,
\end{aligned}
\label{dssq12b}
\end{equation}
which is, again, purely a function of $r$. 
 
Like the $(1,0,n)$ vector fields in  (\ref{Max1}), the vector field $\hat A^{(\phi)}$ is independent of the coordinates on the $S^3$ and, when incorporated in the metric,  is multiplied by a Killing vector.  This means that it reduces to a massless Kaluza-Klein vector field on $\cK$.  However, $\hat A^{(\psi)}$, while also multiplying Killing vectors in the metric, has terms that are independent of $\theta$ as well as terms proportional to $\cos 2 \theta$.  The former are constant multiples of $dv$ and are thus pure gauge.  The latter also depend on $r$ and therefore represent non-trivial profiles for massive vector fields on $\cK$.  The metric of the $(2,1,n)$ family thus produces both {\it massive} and  {\it massless} KK vector fields on $\cK$.

\subsubsectionmod{Geodesics and separability}
\label{ss:Example2-waves}

For the $(2,1,n)$ family of solutions, the massless wave equation and the Hamilton-Jacobi equation for null geodesics are separable only for either vanishing frequency, or for a specific choice of angular modes on $\bbS^3$. Specifically, if one seeks modes of the form   (\ref{SepFns}) then  one finds something very similar to (\ref{SepEqns1})--(\ref{SepEqns2}) except for a single problematic term.  One finds: 
\begin{equation}
\begin{aligned}
\frac{1}{K} \bigg[ & \frac{1}{r} \partial_r \big( r(r^2 +a^2) \,  \partial_r K \big)\bigg]   ~+~ \frac{1}{S} \bigg[ \frac{1}{\sin \theta \cos \theta} \partial_\theta \big( \sin \theta \cos \theta \,  \partial_\theta S \big)-  \Big(\frac{q_1^2}{\sin^2 \theta} + \frac{q_2^2}{\cos^2 \theta}\Big)\,S  \bigg]  \\
&~+~F(\omega, p, q_1,q_2,n; r)  ~-~  \, a^2 b^2 \, (n+1) \, G(\omega, q_1, q_2,n; r, \theta)~=~ 0
\end{aligned}
  \label{SepEqns3}
\end{equation}
where $F(\omega, p, q_1,q_2,n; r)$ is a complicated function of the coordinate $r$ and the mode numbers $\omega, p, q_1, q_2$ and $n$. The function $G(\omega, q_1,q_2,n; r, \theta )$ is given by:
\begin{equation}
G(\omega, q_1,q_2,n; r, \theta ) ~\equiv~  \omega\,  (q_1 +q_2 )\, \frac{ r^{2n}}{(r^2 +a^2)^{n+2}}  \, \cos 2\theta 
  \label{Gfndefn}
\end{equation}
and expresses the failure of separability.  Note that this term vanishes if $\omega =0$  or $q_2 = -q_1$.   Moreover, the function $G$ is strongly peaked at the dimple of $\Lambda$,  (\ref{peak2}), and vanishes rapidly as $r\to 0$ and $r\to \infty$.   

One finds a similar result upon attempting to separate the massless Hamilton-Jacobi equation:
\begin{equation}
 g^{PQ} \,  \frac{ \partial \, S}{\partial z_P} \, \frac{ \partial \, S}{\partial z_Q}   ~=~ 0 \,.
  \label{HJeqn}
\end{equation}
Substituting 
\begin{equation}
S ~=~ K(r) + S(\theta) + \frac{\sqrt{2}}{R_y}E \, u  + \frac{\sqrt{2}}{R_y} \ell_0\, v +  \ell_1 \, \varphi_1 +  \ell_2\, \varphi_2 \,,
  \label{HJsep1}
\end{equation}
one obtains an equation of the form:
\begin{equation}
\begin{aligned}
 \bigg((S'(\theta))^2  ~+~ \frac{ \ell_1^2}{\sin^2 \theta}  ~+~ \frac{\ell_2^2}{\cos^2 \theta}\bigg) ~+~ & \bigg( (r^2 +a^2) \, (K'(r))^2 ~+~ \widehat F(E, \ell_0, \ell_1,\ell_2, n; r)  \bigg) \\ &  ~+~   \,(n+1)\, a^2 b^2 \,\widehat G(E, \ell_1,\ell_2, n; r,\theta)~=~ 0  \,,
\end{aligned}
  \label{HJsep2}
\end{equation}
where
\begin{equation}
\widehat G(E, \ell_1,\ell_2, n; r,\theta) ~\equiv~ E\,(\ell_1+\ell_2)  \,  \frac{ r^{2n}}{(r^2 +a^2)^{n+2}}  \, \cos 2\theta    \,.
  \label{HJsep3}
\end{equation}
This is manifestly the direct parallel of   (\ref{Gfndefn}).  Moreover, if either $E$ vanishes or $\ell_1 +\ell_2$ vanishes, then $\widehat G \equiv 0$ and  Hamilton-Jacobi theory tells us that 
\begin{equation}
 \bigg(g_{\theta\theta }\frac{d \theta}{d \lambda} \bigg)^2  ~+~ \frac{ \ell_1^2}{\sin^2 \theta}  ~+~ \frac{\ell_2^2}{\cos^2 \theta}  
 ~=~ \Sigma^2 \,\cP\,  \bigg(\frac{d \theta}{d \lambda} \bigg)^2  ~+~ \frac{ \ell_1^2}{\sin^2 \theta}  ~+~ \frac{\ell_2^2}{\cos^2 \theta}    
  \label{cons2}
\end{equation}
is a conserved quantity. Note that one has
\begin{equation}
 \Sigma^2 \,\cP  ~=~    \frac{R_y^2}{2} \,(2 a^2 +b^2)\, \bigg(1 -\frac{2 (n+1)  \,a^4\,b^2}{(2 a^2 +b^2)} \, \frac{r^{2n}}{(r^2 +a^2)^{n+2}} \, \sin^2 \theta\, \cos^2 \theta   \bigg)~=~    \frac{R_y^2}{2} \,(2 a^2 +b^2)\, \Lambda^2\,,
  \label{Sig2P}
\end{equation}
where $\Lambda$ is given in (\ref{Lambdadef2}).   Thus (\ref{cons2}) is the analogue of the conserved quantity (\ref{ConKtens}).  However, (\ref{cons2}) is only conserved for $E=0$ or $\ell_1 = -\ell_2$.  

Recall that the momentum modes that underlie our solution depend upon the angles according to (\ref{SSmodes1}), which  now has the form: 
\begin{equation}
\chi_{2,1,n} ~\,\equiv\,~ \coeff{\sqrt{2}}{ R_y}\,   (n+1) \, v ~+~   (\varphi_1 -  \varphi_2)   \,.
\end{equation}
Thus the conservation and separation conditions, $\ell_1 = -\ell_2$ and $q_2 = -q_1$, mean that the geodesic or wave must have the same angular dependence  on the $\bbS^3$ as the underlying momentum modes.

\newpage

\section{Discussion}
\label{sec:Disc}

In this paper we have found that two infinite families of superstratum solutions have quite remarkable integrability properties for null geodesics.
One of the families, the $(1, 0,n)$ family, has a separable massless Klein-Gordon equation and a complete set of conserved quantities for null geodesics.  The other family, the $(2, 1,n)$ family, has a separable massless Klein-Gordon equation and a complete set of conserved quantities only for a constrained set of angular momenta on the $\bbS^3$.  For the $(2, 1,n)$ family (and for the $(2, 0,n)$ family), the failure of separation and failure of conservation  is  sharply localized in the region of the solution where the momentum density is concentrated.
We further found that the metrics of these families of solutions can be reduced to interesting sets of degrees of freedom in $(2+1)$-dimensions. 

The conservation laws  and separability of the massless scalar wave equation for  the $(1, 0,n)$ family means that this solution is readily amenable to detailed scattering calculations.  On the other hand, because of its integrability, this solution is likely to exhibit some quite atypical behavior, particularly when it comes to the spectrum.  

An interesting question to investigate is whether, and in what regime of parameters, a given microstate geometry can capture or trap incoming particles. It was recently argued that given any supersymmetric microstate geometry in six dimensions, there should exist a stably-trapped null geodesic passing through every point of the spacetime~\cite{Eperon:2016cdd}. These null geodesics have tangent vector $\partial/\partial u$ in our notation, so correspond to massless particles moving purely in the $y$ direction. In the geodesic approximation, massless particles following such geodesics do not fall (deeper) into the throat. The main heuristic argument of \cite{Eperon:2016cdd} considers such a particle that is coupled to gravitational radiation and other massless  fields, such that the probe gradually radiates some of its energy into these other fields, thus evolving to follow geodesics of progressively lower energy. In this way a massless particle can, slowly, descend the throat.

However, within the geodesic problem, one can ask whether an infalling particle with non-zero radial momentum, falling from outside the throat, can be deflected non-trivially from the region of the metric at the bottom of the throat and, through this deflection, remain in the throat for arbitrarily long periods of time, as seen from infinity.  If the ``radial kinetic energy,'' $\frac{1}{2} v_r^2$, is the only kinetic energy term that appears in a particular conservation law, then, just as in any orbit problem, an infalling particle falling from outside the throat will simply rebound and escape: it will not be captured and trapped deep within the throat.  For such capture to occur, particles and waves must be able to scatter ``radial kinetic energy,'' $\frac{1}{2} v_r^2$, into ``angular kinetic energy.'' A complete set of conserved quantities is thus largely antithetical to such behavior, although the conserved quantity,  (\ref{ConKtens}),  depends on  both $r$ and $\theta$ via (\ref{Lambdadef1}) and so, in principle, it is possible that changing $\theta$ along the trajectory can result in the  loss of some radial kinetic energy.   However, in practice, in our solutions the $\theta$ dependence dies out extremely rapidly for large $r$, and so changing $\theta$ will have only a minimal effect upon the return of the particle to large distances.  

If a given solution does allow a significant deflection of radial momentum to angular momentum, the angular motion can potentially prevent the particle from escaping the throat for a long time. Our results imply that the $(1, 0,n)$ family of solutions is likely the wrong place to look for such behavior.  However, the conservation laws present for the $(1, 0,n)$ family are not present for generic $(k, m,n)$ superstrata, so such solutions should allow the scattering of $v_r$ into angular motion, and it would be interesting to investigate whether this can lead to geodesics that describe infalling particles with non-zero radial momentum, falling from outside the throat, becoming trapped deep inside the throat for long periods of time. 
One way to study this behavior analytically could be to use the $(2, 1,n)$ family (for which geodesics/waves with $\ell_1 + \ell_2=0$ or $q_1+ q_2=0$ are integrable)  and examine perturbatively how waves are scattered into angular directions for small $j_L=\ell_1 + \ell_2$ or small $q_1+ q_2$.  It would be particularly interesting to investigate the timescale associated with this trapping.  This will depend on the depth of the throat, on how radial motion is converted into angular motion,  and on whether or not the trapping is chaotic.   It might be that the only way in which a particle can return to large distances after having scattered off the microstate structure once is to scatter off it again in exactly the right manner as to restore enough radial kinetic energy.  This could lead to extremely large return times -- a desirable feature if one is to construct microstate geometries that describe typical black hole states.

Finally, the results presented here underline the fact that the $(1,0,n)$ and $(2,1,n)$  families have remarkably stringent constraints on their structure that suggests that the full six-dimensional solutions might be written in a form that would come from a consistent truncation to $(2+1)$-dimensions. In particular, the Fourier expansions of the $(2+1)$-dimensional fields may only have non-trivial dependence on one variable, $r$. If such a structure indeed exists, it would be very interesting to investigate the existence or otherwise of a consistent truncation containing these solutions. In doing so, an interesting question will be to determine whether or not any consistent truncation ansatz will require some form of coiffuring to be built in. If a consistent truncation involving the modes of the tensor gauge fields exists, then this could provide a powerful new route for the construction non-supersymmetric solutions building on~\cite{Bossard:2014ola,Bena:2015drs,Bena:2016dbw}: It would reduce the non-linear, non-BPS supergravity dynamics in the system of~\cite{Bossard:2014ola,Bena:2016dbw} from functions of two variables in six dimensions to the far more tractable problem of functions of one variable in $(2+1)$-dimensions. These questions are currently under investigation.

\vspace{3mm}

\section*{Acknowledgments}
\vspace{-0.5mm}
NPW is very grateful to the IPhT of CEA-Saclay, and IB and DT to the {\it Centro de Ciencias de Benasque}, for hospitality during this project.  We would like to thank Stefano Giusto, Mariana Gra\~na, Monica Guic\u a, Stefanos Katmadas, Emil Martinec, Rodolfo Russo, Masaki Shigemori and Charles Strickland-Constable for interesting discussions. 
The work of IB and DT was supported by the ANR grant Black-dS-String. 
The work of NPW was supported in part by the DOE grant DE-SC0011687.
The work of DT was also supported by a CEA Enhanced Eurotalents Fellowship.

\newpage

\begin{adjustwidth}{-1mm}{-1mm} 
\bibliographystyle{utphys}      
\bibliography{microstates}       

\end{adjustwidth}


\end{document}